\documentclass[preprints,article,accept,pdftex,moreauthors]{Definitions/mdpi} 
\firstpage{1} 
\pubvolume{1}
\issuenum{1}
\articlenumber{0}
\pubyear{2025}
\copyrightyear{2025}
\datereceived{ } 
\daterevised{ } 
\dateaccepted{ } 
\datepublished{ } 
\hreflink{https://doi.org/} 

\usepackage{comment}
\usepackage{amssymb}
\usepackage{ulem}

\Title{Strangeon Matter: from Stars to Nuggets}

\TitleCitation{Strangeon Matter: from stars to nuggets}

\Author{Haoyang Qi $^{1}$\orcidA{} and Renxin Xu $^{1}$\orcidB{}}

\AuthorNames{Haoyang Qi and Renxin Xu}

\isAPAStyle{%
       \AuthorCitation{Qi, H., \& Xu, R.}
         }{%
        \isChicagoStyle{%
        \AuthorCitation{Haoyang Qi, and Renxin Xu.}
        }{
        \AuthorCitation{Qi, H.; Xu, R.}
        }
}

\address{%
$^{1}$ \quad School of Physics, Peking University, Beijing 100871, China; qhy021@pku.edu.cn, r.x.xu@pku.edu.cn}

\abstract{The fact that strange sea quarks are abundant in the nucleons, but with zero net strangeness, is of great importance for understanding the nature of matter condensed by the strong interaction, particularly in the context of the ``gigantic nucleus'' created after the gravitational collapse of an evolved massive star.
We hypothesize that the basic unit of bulk strong matter with the approximately light-flavored symmetry of valence quarks is ``strangeon'', which is the counterpart of the nucleon found in atomic nuclei.
In addition to strangeon stars (SnSs) with large baryon number of $A\approx 10^{57}$, strange nuggets (SnNs) with $A\gtrsim 10^{10}$ could also exist in the Universe.
Both the SnS and the SnN are explained, with attention to the evidence from observation and detection.}


\begin{document}



\section{Introduction}

The motivation of {\em Physics} is to understand the material world. Phenomenologically, physics aims to reveal the fundamental building blocks of matter and the interactions that bind them together, allowing us to know the nature of matter.
The normal material of atoms/ion-electrons condensed by electromagnetic (or simply electric) force was the only form of matter known until the discovery of Ernest Rutherford in 1911~\citep{massspectrum2}.
At that time, Geiger and Marsden, under the guidance of Rutherford, shot $\alpha$ particles towards a stationary gold target and observed particles larger than 1/10000 scattered back at a deflection angle greater than 90 degrees, which was much higher than the prediction given by Thomson's ``plum-pudding'' atomic model. From this result, Rutherford concluded that the vast majority of the atomic mass and all positive charges are concentrated in a region of about a few femtometers' radius at its center, which is called the atomic nucleus.
In 1919, Rutherford shot nitrogen nuclei with $\alpha$ particles and obtained protons. In 1932, Chadwick shot beryllium nuclei with $\alpha$ particles and analyzed the presence of neutrons in the product, confirmed Rutherford's conjecture from the charge to mass ratio of atomic nuclei in 1920.
These results revealed a novel matter of nucleons, condensed by what is now known as the fundamental ``strong interaction''.
For simplicity, we will refer to the former as $electric$ matter, and the latter as $strong$ matter.

\underline{Can strong matter manifest only in the form of atomic nuclei?}
The answer was for long time considered to be ``yes'', even pondering the remnants of evolved massive stars, i.e., the ``gigantic nucleus'' initiated by Lev Landau in 1932~\citep{landau_theory_1932}, the so-called ``neutron sphere''.
It is noteworthy that the key point of Landau's idea of a ``neutron star'' is to neutralize the constituent elements --- creating an almost charge-neutral state for nucleons so that a relatively small amount of electrons
would contribute negligible kinetic energy.
As an exemplar of fermions following the Fermi-Dirac statistics, an electron must populate a higher energy level if the lower levels are occupied, as is in the case with atoms that have many electrons. For a number density of $n$, the mean separation between electrons $\Delta x$ is approximately $n\sim \Delta x^{-3}$. From the Heisenberg relation, $\Delta x\cdot \Delta p\sim \hbar$, we can obtain the Fermi momentum of an electron $p_{\mathrm F}\simeq \hbar n^{1/3}$. So when the number density of electrons $n$ is low, the Fermi kinetic energy contained in the electrons is also relatively low. The number density of electrons in a neutral system can only be lowered by reducing the charge-to-mass ratio of quarks or nucleons. Therefore, a small charge-to-mass ratio, $R_{\textbf{em}}$, is preferred in terms of kinetic energy. We would therefore think that a gigantic nucleus should be a neutron star because $R_{\textbf{em}}$(neutron) = 0, whereas $R_{\textbf{em}}$(proton) = 1. In this sense, Landau's neutron star is, in fact, a neutral star~\citep{xu2023}. Besides the electron gas, nonetheless, there are also nucleons in the system, and the so-called symmetry energy emerges dramatically for a neutron-rich gigantic nucleus. But Landau was unaware of the symmetry energy in 1932. That is like ``{\em Press down the gourd, but the ladle pops up}'' in a Chinese proverb. The symmetry energy of nuclei was first addressed by Weizs{\"a}cker in 1935~\cite{Weizsacker1935}. The symmetry energy is higher when there is a significant difference in the number of neutrons and protons in a nucleus, that is, the state now known as isospin asymmetry or two-flavor asymmetry.

\begin{figure}[H]
\isPreprints{\centering}{} 
\includegraphics[width=8cm]{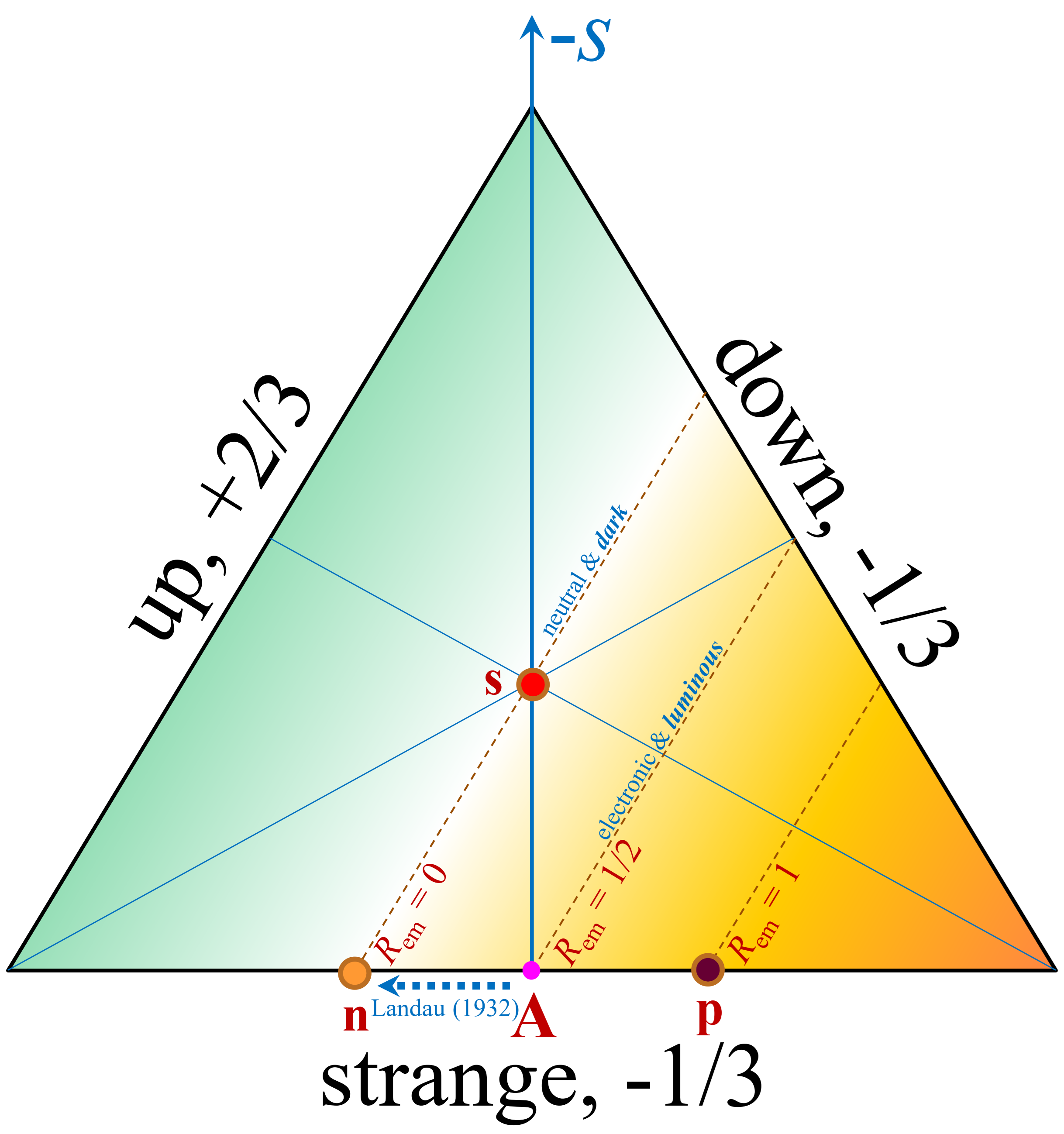}
\caption{%
Triangle of light-quark flavors ($u$, $d$, and $s$)~\citep{triangle}.
A point inside the triangle defines a state with a specific ratio of up, down and strange quarks, as measured by the point's distance from each of the triangle's edges.
The charge-to-mass ratio of the baryonic component, $R_{\textbf{em}}$, is illustrated by the depth of the color.
Atomic nuclei are at point ``A'', pure neutron matter is at point ``n'', pure proton matter is at point ``p'', strange quark matter and strangeon matter (SnM) are at point ``s''.\label{tri}}
\end{figure}

However, more than 90-years later, especially after the establishment of the standard model of particle physics, \underline{can an alternative, and possibly better way, to form a charge-neutral system be} \underline{devised today?}
Yes, the strangeness matters!
In comparison with the low-energy system of atom  (nucleus and electrons), there are abundant quark-pairs inside the high-energy system of nucleon (e.g., the complex proton\footnote{%
This is also simply due to the Heisenberg relation. An atom is a state bound by the electromagnetic force, while a proton is a state bound by the strong force. This results in a relatively large atom but a smaller proton. According to the Heisenberg relation, the typical kinematic energy of an electron in a hydrogen atom is $\sim 10$~eV, whereas that of a quark in a proton is $E_{\rm scale}\sim 300$~MeV~\cite{strongM}. The fact that the bare masses of all the light-flavor quarks are $<E_{\rm scale}$ (but the bare masses of all the heavy-flavor quarks are $\gg E_{\rm scale}$) means that abundant strange-quark pairs do exist in a nucleon even without net strangeness. In fact, like any other bound states (e.g., hadrons), the proton is an object of unspeakable complexity: filled with the dance of short-lived $uds$-quarks (i.e., the sea quarks) as well as gluons! Nevertheless, the time-averaged number of net $u$-quarks is {\bf 2} (i.e., the number of $u$-quarks minus that of $\Bar{u}$-quarks), and the number of net $d$-quarks is {\bf 1}, but the number of net $s$-quarks is {\bf 0}. We can therefore conclude that the valence quarks in a proton are \{$uud$\} (or \{$u^2d^1s^0$\}).
}). %
It is then convenient to consider the charge-neutrality using the triangle illustrated in Figure~\ref{tri}.
Given that the charges of $d$- and $s$-quarks are equal, the depth of the color of the image in Figure~\ref{tri} can be used to represent the charge-to-mass ratio, $R_{\textbf{em}}$.
When a massive star reaches the end of its life, its core loses energy from nuclear reactions and undergoes gravitational collapse. Under extreme compression, the density increases to that of nuclear matter. At this point, the Fermi kinetic energy of electrons can reach about 400\,MeV, higher than the energy required for the inverse $\beta$ reaction and even higher than the rest mass of $s$ quarks. So, the compressed core of the star undergoes a neutralization transition to reduce the enormous electron Fermi energy.
Landau (1932) only superficially predicted the transition of atomic nuclei from point ``A''  to a gigantic one at point ``n'', but allowing only for two asymmetric flavors of valence quarks.
However, the most beautiful (i.e., symmetric) place in the triangle should be at the point ``s'', with both charge neutrality and 3-flavored symmetry. By reducing the charge-to-mass ratio, the electron Fermi energy of the system can be lowered without encountering the problem of increasing symmetry energy.
We refer to the matter at the point ``s''  as ``strange matter'', which contains almost equal numbers of $u$, $d$ and $s$ valence quarks.

The strange matter was already proposed in the 1970s. Afterwards, the models of strange matter continue to evolve. Bodmer hypothesized that strange quark matter would be produced when atomic nuclei collapse and squeeze together~\citep{bodmer_collapsed_1971}.
Bodmer's main focus is on medium-sized nuclei with baryon number O(10). After collapse, the nucleons disintegrate, forming quark clusters. Bodmer discussed the properties of collapsed nuclei and the possibility of their widespread existence in the universe. Meanwhile, Bodmer proposed that although existing ordinary atomic nuclei are in a metastable state, the average collapse time scale is as long as $10^{31}$\,s.
Then in the well-known MIT bag quark matter model~\citep{johnson_mit_1975}, strange quarks were also taken into account. The main idea of the MIT bag model is to introduce a non-perturbative QCD vacuum in which quarks are confined. This vacuum, which separates it from the real vacuum with no deconfined quarks, has an additional constant vacuum energy, commonly known as the ``bag constant'' $B$.
Witten further proposed that strange matter is the true ground state of the strong interaction, and explored the consequences of the emergence of strange quark matter in the early universe~\citep{witten_cosmic_1984}. In addition, from the perspective of high quark chemical potential and Fermi energy balance, Witten also believed that strange quark matter composed of almost equal amounts of $u$, $d$, and $s$ quarks has lower energy than other ratios such as 2:3:1 $u$, $d$, and $s$ quarks.
Farhi and Jaffe~\citep{farhi_strange_1984} discussed in more detail the properties of small baryon number ($A<100$), intermediate baryon number ($100<A<10^7$), and bulk strange quark matter under the modified MIT theoretical framework including O($\alpha_{\rm c}$) order correction of perturbative QCD and correction from $s$ quark mass $m_s$.
Alcock et al.~\citep{alcock_strange_1986} calculated the macroscopic parameters such as mass, radius, moment of inertia of strange quark stars; discussed some observable phenomena such as the absence of thermal X-ray photons and the influence of the possible crust; and speculated on some ways of formation of the strange quark stars such as weak-interaction transition from two-flavor quark matter, clustering and deconfining from $\Lambda$ hyperons that transitioned from the protons and neutrons, crossing the strangenization barrier with temperature at the birth of the neutron star, high-energy neutrino induction, and external strange particle incidence and induction.

Although the strange matter is called ``strange'', it is not unusual at all. Because the size of the nucleus is about 1\,fm, the momentum of quarks inside nucleons obtained from Heisenberg's uncertainty relation is approximately 200\,MeV$/c$. Meanwhile, due to the presence of strong interactions, many high-order Feynman diagrams play a role in the interactions between quarks. So, strange quark $s$ (and anti-strange quark $\overline{s}$) components will definitely appear in the quark sea. In fact, people have already measured the composition of strange sea quarks in nuclear physics experiments.
Note that the momentum fraction ratio carried by the strange sea quarks is from $\sim$ 5\% to 10\%~\citep{ellis_strangeness_2001,chang_flavor_2014}. This result reflects the breaking of flavor SU(3) symmetry, but at the same time suggests that flavor SU(3) symmetry may be a relatively good symmetry at several times the nuclear saturation density inside the compact stars.
The specific process of exciting $s$ quarks from the Dirac Sea to the Fermi Sea remains uncertain. Alcock~\citep{alcock_strange_1986} mentioned some possible pathways, such as direct vacuum polarization of quark-anti-quark pair $s \overline s$, weak interaction transition from $u$ or $d$ quarks, or excitation by high-energy incident particles or strange particles.
Charm quarks or heavier ones also contribute to the quark sea, but due to their bare mass greater than 1\,GeV, this contribution is minimal. We should consider lighter three flavor quarks ($u, d, s$), which play an important role in the properties of dense matter that is under zero pressure and is condensed by the strong interaction.

\underline{Are quarks truly the fundamental building blocks of strange matter?}
Yes, strange matter is favored by both the charge neutrality and the flavor symmetry of quarks, but the puzzling non-perturbative nature of the quantum chromo-mechanism (i.e., NQCD) does work in this case.
There would then be two significant consequences of NQCD: quark-clustering (grouped in position space) and core-repulsing (hard core).
Both these two are phenomenologically understood for atomic nuclei, but can hardly be explained by NQCD even today.
In fact, there have been many theoretical attempts in this problem.
Glendenning~\citep{glendenning_first-order_1992} proposed that there may be a hadron-quark first-order phase transition in the compact star; in the high-density and high-pressure region of the core, matter exists in the form of deconfined quarks.
Hatsuda et al.~\citep{hatsuda_qcd_1994} applied the Nambu-Jona-Lasinio (NJL) model~\citep{nambu_dynamical_1961a,nambu_dynamical_1961b}, a chiral effective field theory, to the calculation of the internal structure of neutron stars.
Color superconductivity, neutron superfluid, color-flavor locking, and pinning phenomena in the compact star were subsequently considered~\citep{negele_neutron_1973,barrois_superconducting_1977,alford_qcd_1998,alford_color-flavor_1999,alford_crystalline_2001,alford_minimal_2001}.
Some people have attempted to introduce hyperon components into compact stars~\citep{schaffner_hyperon-rich_1996,prakash_composition_1997}, but this often softens the equation of state, and causes a smaller maximum mass, which is not consistent with existing observations such as PSR J1614-2230~\citep{J1614-2230}. This problem is called the ``hyperon puzzle'', and some studies attempt to explain this issue by limiting the abundance of hyperons or altering the description of interactions~\citep{weissenborn_hyperons_2012,lonardoni_hyperon_2015,burgio_neutron_2021}.
A summary of various nuclear states and simple MIT bag quark states, and the influence of equations of state on the structure and evolution of compact stars can be found in~\citep{lattimer_neutron_2001,lattimer_physics_2004}.
A detailed review of the NJL model, color superconductivity, color-flavor locking, and their existence in the compact stars can be found in~\citep{buballa_njl-model_2005}.
In recent years, a review of the structure and equation of state of compact stars can be referred to~\citep{chu_quark_2014,ozel_masses_2016,weber_phases_2019}.

Nevertheless, as an analogy of nucleons in point ``A'', one could also expect a nucleon-like unit in point ``s''~\cite{strangeon}.
We name this unit ``strangeon'',\footnote{%
It should be noted that strangeon differs from hyperon in at least two ways.
The hyperon's baryon number, $B$, is 1, whereas a strangeon, as a multi-quark bound state, is not simply a baryon. Previous studies have discussed multiquark states such as hidden-charm pentaquark states, hadronic molecules, and exotic hadrons consisting of undetermined number of quarks and gluons, both theoretically and experimentally based on data~\citep{chen_hidden-charm_2016,guo_hadronic_2018,wu_deepquark_2025,husken_brief_2025}. A strangeon would behave like an exotic hadron with baryon number $B > 1$, likely $B = 6$ (just as Michel's quark alpha, Q$_\alpha$~\citep{michel_quark_1988,Bayesian}, it can satisfy color singlet, spin symmetry, and flavor symmetry simultaneously) and would be even larger inside massive strangeon stars with masses approaching the mass limit.
Additionally, the interaction between two $\Lambda$-particles could be attractive without a hard-core. Many previous studies have shown this point~\citep{beane_evidence_2011,inoue_bound_2011,mathur_strongly_2023,wu_masses_2024}. However, strangeons are supposed to similar to nucleons with hard-core. The discussion on the hard core between nucleons can be referred to~\citep{wilczek_particle_2007,sargsian_hole_2024}. A short-distance repulsive interaction between strangeons would be naturally expected, although the physical reason for this is still unclear in non-perturbative quantum chromodynamics (QCD).
The hard-core and non-relativistic nature of massive strangeons could both result in an equation of state that is stiff in a model-independent way.} %
and are explaining strangeon matter (SnM) in the following sections.
It is worth noting that the main difference between SnM (including strangeon nuggets) and strange quark matter (including Witten's strangelets) is their basic unit. The basic units of SnM are hadron-like strangeons, but the basic units of strange quark matter are deconfined quarks. The strangeon matter is solid at zero temperature because the basic units, strangeons, are very massive, act like classical particles, and form a classic solid. The quark matter is usually liquid, due to the small mass of quarks and the formation of color-flavor locking and color superconducting quantum states.

\section{An evolving trend during gravitational contraction}

Let us begin with a gravitational system (mass $M$ and radius $R$) that has spherical symmetry.
It seems that the system would be most stable when $R\rightarrow 0$ (i.e., becoming a black hole), since its gravitational energy is the minimum, $E_{\mathrm g}\simeq -GM^2/R\rightarrow-\infty$, in this case.
However, there is an internal energy of matter, and the system could temporarily reach a stable state during gravitational contraction if the internal energy is comparable to $-E_{\mathrm g}$.
This is why, besides black holes, both main-sequence and compact stars do exist in the Universe.

The total energy of the system reads
\begin{equation}
E(M)=\epsilon\frac{M}{m_{\mathrm u}}+E_{\mathrm g}
\simeq \epsilon\frac{M}{m_{\mathrm u}}-\frac{GM^2}{R}
=(\epsilon-\frac{GM{m_{\mathrm u}}}{R})\frac{M}{m_{\mathrm u}},
\end{equation}
where $\epsilon$ is the internal energy of matter per baryon, $m_{\mathrm u}c^2=931.5$\,MeV $\sim 1$ GeV is the atomic mass unit, $M/m_{\mathrm u}$ represents the total baryon number, $A$. When the system reaches an equilibrium, its internal energy is comparable to the absolute value of gravitational energy, one then has
\begin{equation}\label{eq:radius}
R\approx\frac{GMm_{\mathrm u}}{\epsilon}=\frac{GM_\odot}{c^2}\frac{M}{M_\odot}\frac{m_{\mathrm u}c^2}{\epsilon}
\simeq 1.5\,\mathrm {km}\frac{M}{M_\odot}\frac{1~{\rm GeV}}{\epsilon}.
\end{equation}
In other words,
\begin{equation}\label{eq:internalenergy}
\epsilon
\simeq 1~{\rm GeV}\frac{M}{M_\odot}\frac{1.5\,\mathrm{km}}{R}.
\end{equation}

\begin{figure}[H]
\isPreprints{\centering}{} 
\includegraphics[width=12 cm]{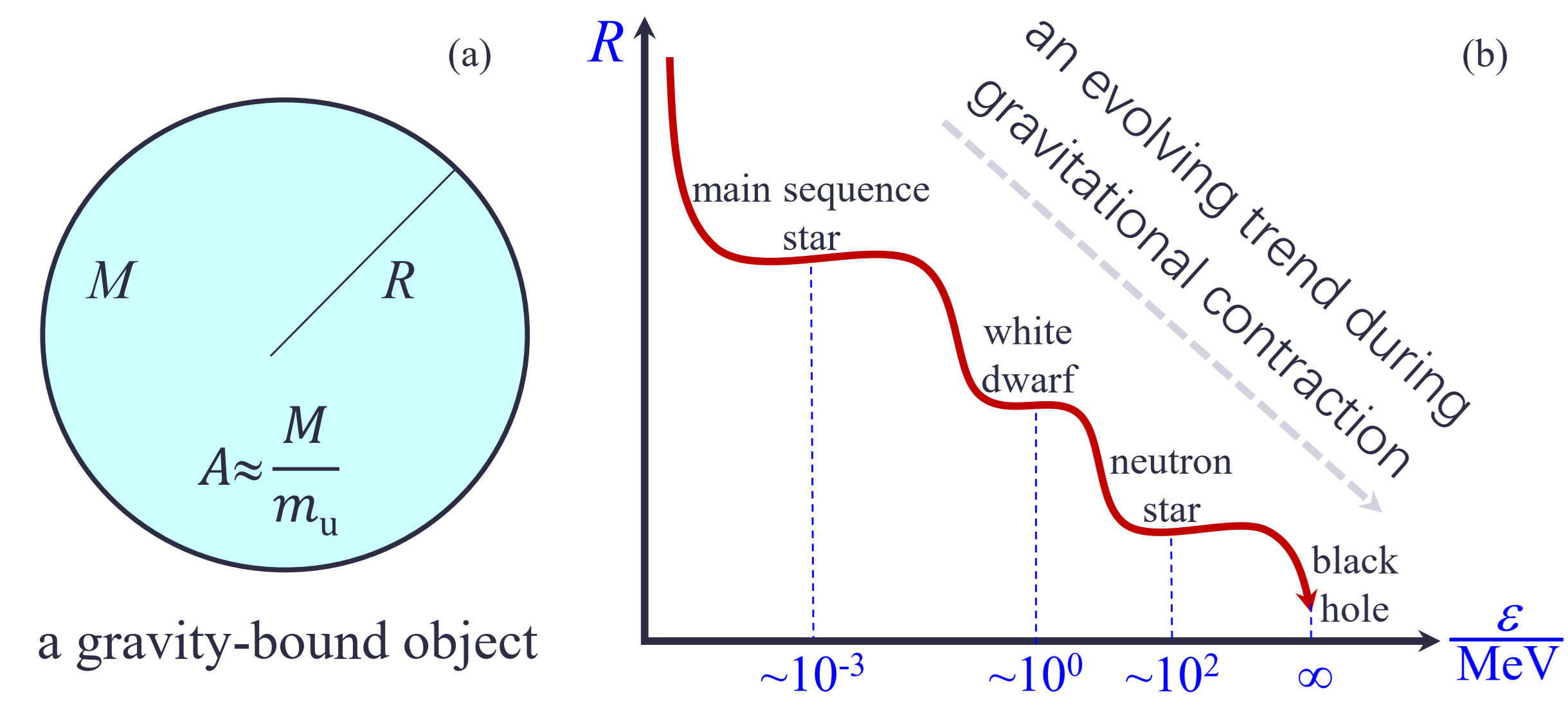}
\caption{The three equilibrium steps before an object with a high initial mass collapses into a black hole. The three steps are: first, main-sequence stars, which are supported by thermal gas pressure and nuclear power; second, white dwarfs, which are supported by fermionic degeneracy pressure supplied by the electron gas; and third, neutron stars, which are supported by the strong interaction (e.g. hard core) between baryons in the non-perturbative regime. These are ordered by radius from large to small and by internal energy per particle from low to high.\label{stars}}
\end{figure}

As a system with negative heat capacity, the kinetic internal energy, $\epsilon$, increases as the gravity-bound object releases more energy (i.e. as the radius decreases).
This fact results in the emergence of various objects, from main-sequence stars ($\epsilon\sim$ keV) to white dwarfs ($\epsilon\sim$ MeV) and neutron stars ($\epsilon\sim 10^2$ MeV), and eventually black holes ($\epsilon\rightarrow+\infty$), as illustrated in Figure~\ref{stars}.
Certainly, we assume that the initial mass of the system is high enough to go toward a black hole, otherwise it would stop at a step of a white dwarf or a neutron star.

In Figure~\ref{stars}, the first step represents the main-sequence stars. The nucleosynthesis process is ongoing in the core of the main-sequence stars, maintaining a temperature gradient throughout the stars that allows the thermal pressure of the ionized gas to be strong enough to compete with gravity. The thermal kinetic energy of a particle in the core of the stars is $10^{-3}$\,MeV (corresponds to the nucleosynthesis temperature, $10^7$\,K), the radius of a main-sequence star is then $R\simeq 10^6\,{\rm km} \cdot M/M_\odot$ from Equation~\ref{eq:radius}, for the Sun as an example.

The second step represents the white dwarfs. When low-mass stars (about $0.5\,M_\odot\sim8\,M_\odot$) run out of their nuclear energy, their cores will undergo gravitational collapse and become white dwarfs. When the electron gas reaches high density to become a relativistic degenerate Fermi gas, the equilibrium between the star's self-gravity and the degeneracy pressure will become unstable and therefore there exists a maximum mass $\sim1.44\,M_\odot$, for white dwarfs, called the Chandrasekhar limit~\citep{Chandrasekhar}. So, the internal energy in a white dwarf is comparable to the rest energy of the electrons, 0.5\,MeV. The radius of a white dwarf is then $R\simeq 10^3\,{\rm km}$ from Equation~\ref{eq:radius}.

The third step represents the ``neutron stars''\footnote{Although these objects are commonly referred to as neutron stars in the literature, they do not necessarily consist of neutrons.}, which are manifested in the form of pulsar-like compact stars. A typical observed pulsar has $\sim1.4\,M_\odot$ mass and $\sim10\,$km radius, so the internal energy\footnote{%
This energy is commonly referred to as the degenerate energy of fermions in current textbooks, whether they are neutrons or quarks~\citep{haensel_neutron_2007}.
However, a conceptional mistake could be made here by following the well-trodden path of interpreting the nature of white dwarfs as degenerate electrons, since the interaction energy with a hard-core potential resulting from the non-perturbative QCD works well for the gravitational equilibrium of stable stars too.
From this perspective, we would expect the stabilization mechanism against gravity, which plays a key role in the equation of state of supranuclear matter, to be revised in the era of multimessenger astronomy.
} %
of a particle is $\epsilon\simeq10^2\,$MeV according to Equation~\ref{eq:internalenergy}.
Within this energy range, quantum chromodynamics (QCD) is non-perturbative, the strong interaction can not be negligible but hard to calculate~\citep{densematter}. The basic unit of neutron stars may not be deconfined quarks but be hadron-like particles. In addition, the energy scale is high enough to excite strange quarks as valence ones, and the matter in the neutron star may have non-zero strangeness.
As Fig.~\ref{tri} explains, the color-confined, hadron-like strange particle is called a ``strangeon'', and ``neutron stars'' may actually be strangeon stars (SnSs) in this sense.

When the energy scale reaches $\gtrsim$ 1\,GeV, a (1\,$M_\odot$)-star would contract to its Schwarzschild radius from Equation~\ref{eq:internalenergy}.
This hints that itinerate quarks could hardly appear in ``neutron stars'' because the deconfined quarks may only emerge at the energy scale $\epsilon\gtrsim1\,$GeV, except for objects with masses approaching the maximum mass (e.g., hybrid SnS that contains a strange quark core and a strangeon crust~\cite{hybSnS}).
When all other forms of energy cannot compete with gravitational energy, the internal energy of a particle tends towards infinity and the star inevitably collapses to form a black hole.

\section{Strangeon stars} \label{model}

From the analogy that non-perturbative strong interaction makes quarks form nucleons, we conjecture that the base unit of cold dense matter inside the pulsar is not quarks, but strangeons.
This model was proposed in 2003~\citep{strangeon}, and is constantly evolving, e.g.~\citep{EoS,magnetic,quake,Bayesian}.

\begin{figure}[H]
\isPreprints{\centering}{} 
\includegraphics[width=10.5 cm]{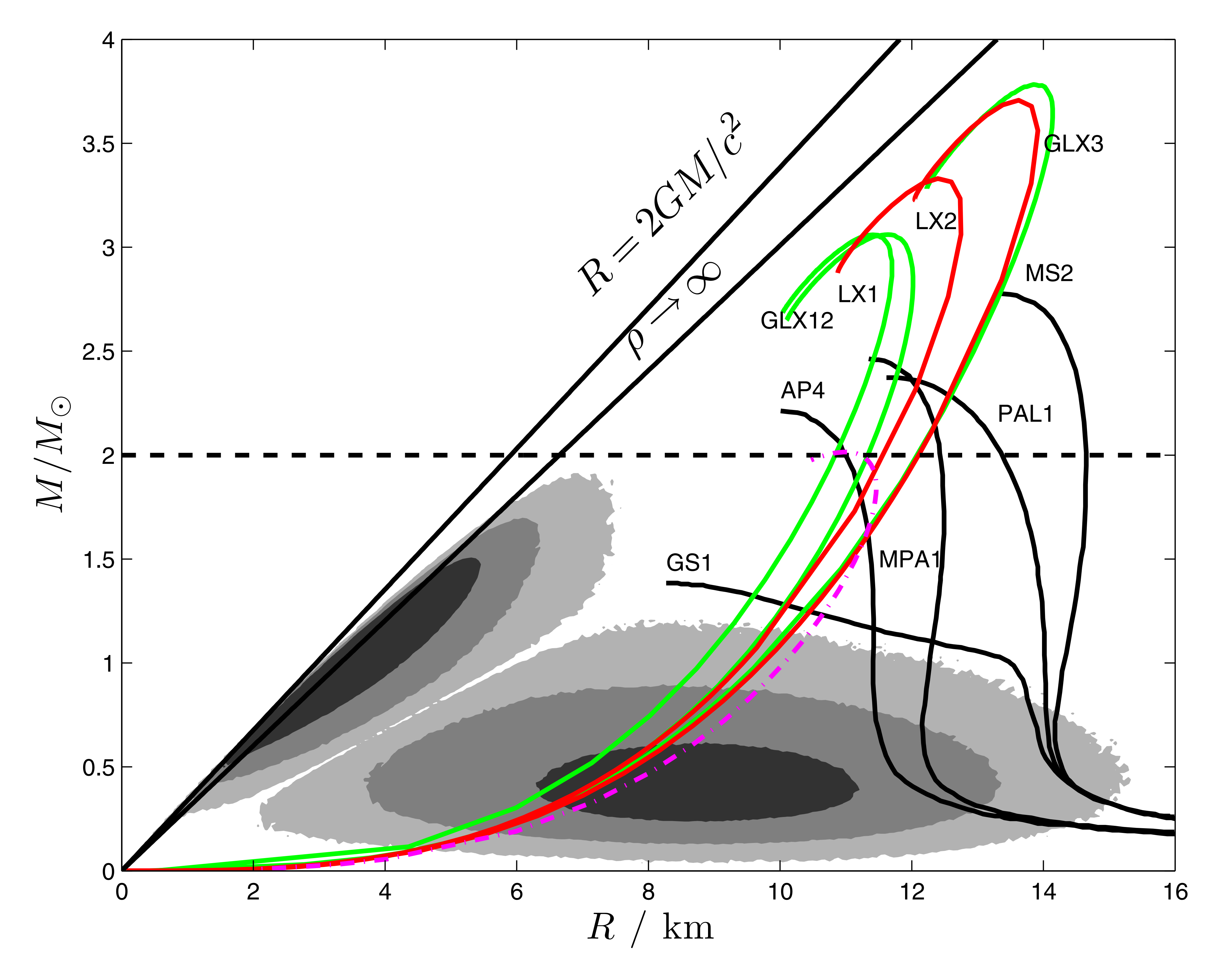}
\caption{The mass-radius curve calculated using Tolman-Oppenheimer-Volkoff equation and equations of state, the theoretical limits from general relativity, and the observational constraints from 4U 1746-37.
This figure is from~\citep{PREburst}. 
The green lines and red lines are from the strangeon-star equations of state with different parameters, the purple dotted-dashed line is from the MIT bag equation of state, the black curves are from the neutron-star equations of state. The black dashed line is $M=2M_\odot$, due to the presence of PSR J0740+6620, the highest point of the MR curve should be higher than this line. The gray regions from deep to shallow are $1\sigma$, $2\sigma$, $3\sigma$ mass-radius confidence regions of 4U 1746-37 based on the assumption that the touchdown flux and emission area were partially obscured by the accretion disk.
It can be seen that the MIT bag quark star and strangeon star models better conform to 4U 1746-37 observation than neutron stars, while the strangeon star model has a larger maximum mass and is more suitable for observing large-mass neutron stars like PSR J0740+6620.}\label{MR}
\end{figure}

There are a lot of mass-radius observations to support the SnS model. The photospheric radius expansion bursts of 4U 1746-37 constrain the mass-radius range of low-mass pulsars, and strangeon-star equation of state~\citep{EoS} meets the constraint better than neutron-star equations of state~\citep{PREburst}, see Figure~\ref{MR}. The discovery of massive neutron stars with mass $M\gtrsim2M_\odot$ such as PSR J1614-2230~\citep{J1614-2230}, PSR J0740+6620~\citep{J0740+6620}, and PSR J0952-0607~\citep{J0952-0607} supports a stiffer equation of state. As seen in Figure \ref{MR}, SnM is stiffer and can reach a $3M_\odot$ maximum mass.

\begin{figure}[H]
\isPreprints{\centering}{} 
\includegraphics[width=8cm]{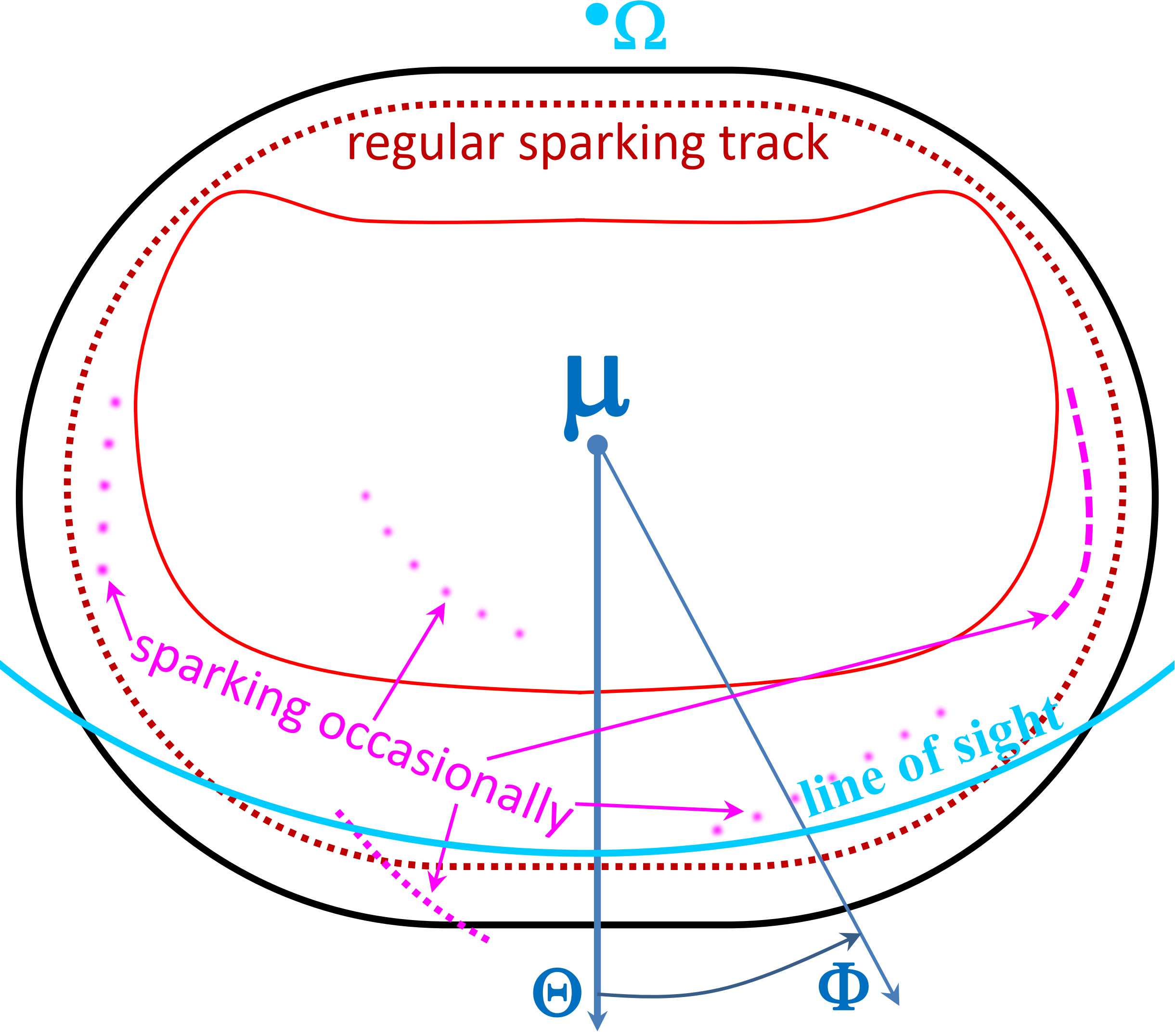}
\caption{
Schematic diagram of the highly linear-polarized spark signal at certain pulse phases caused by zits in the polar cap of the pulsar. The point \textmu is the magnetic pole, the black circle represents the boundary of the polar cap. Some zits on the line of sight spark occasionally and generate a coherent linear-polarized subpulse. 
This figure refers to~\citep{zits1,zits2}.\label{zit}}
\end{figure}

Furthermore, the SnM state can help us to understand subpulses (referring to a short signal structure within a single pulse cycle) in the observed single pulse signals. Because the base unit of SnM are non-relativistic strangeons, the strong interaction between strangeons is strong enough to make the matter solid at low temperature.
Due to the large rest mass of the strangeon, its de Broglie wavelength is relatively small, so it behaves more like a classical particle. Taking an example of electromagnetic matter, when inert gases are cooled at standard atmospheric pressure, only helium will exhibit superfluidity, and heavier elements will combine to form classical solids. Assuming that the Lennard-Jones potential is used to describe the interaction between strangeons, Bayesian analysis of observations shows that the potential well depth of the interaction potential between adjacent strangeons can reach over 10\,MeV~\citep{Bayesian}. When the temperature of strangeon matter is much lower than 10 MeV (which is met by most of neutron stars), we believe that strangeon matter will also combine into classical solids.
This solid SnM is stiffer than a neutron star's solid crust because the former is combined by strong interaction, but the latter is mainly combined by electromagnetic interaction. So, an SnS could support a higher and longer lasting ``zit'' (little hill) than a neutron star's crust could support. These zits may be the spark origin of some highly linear-polarized subpulses~\citep{zits0,zits1,zits2}, see Figure~\ref{zit}.

\section{Strangeon nuggets}

\begin{figure}[H]
\isPreprints{\centering}{} 
\includegraphics[width=12.8 cm]{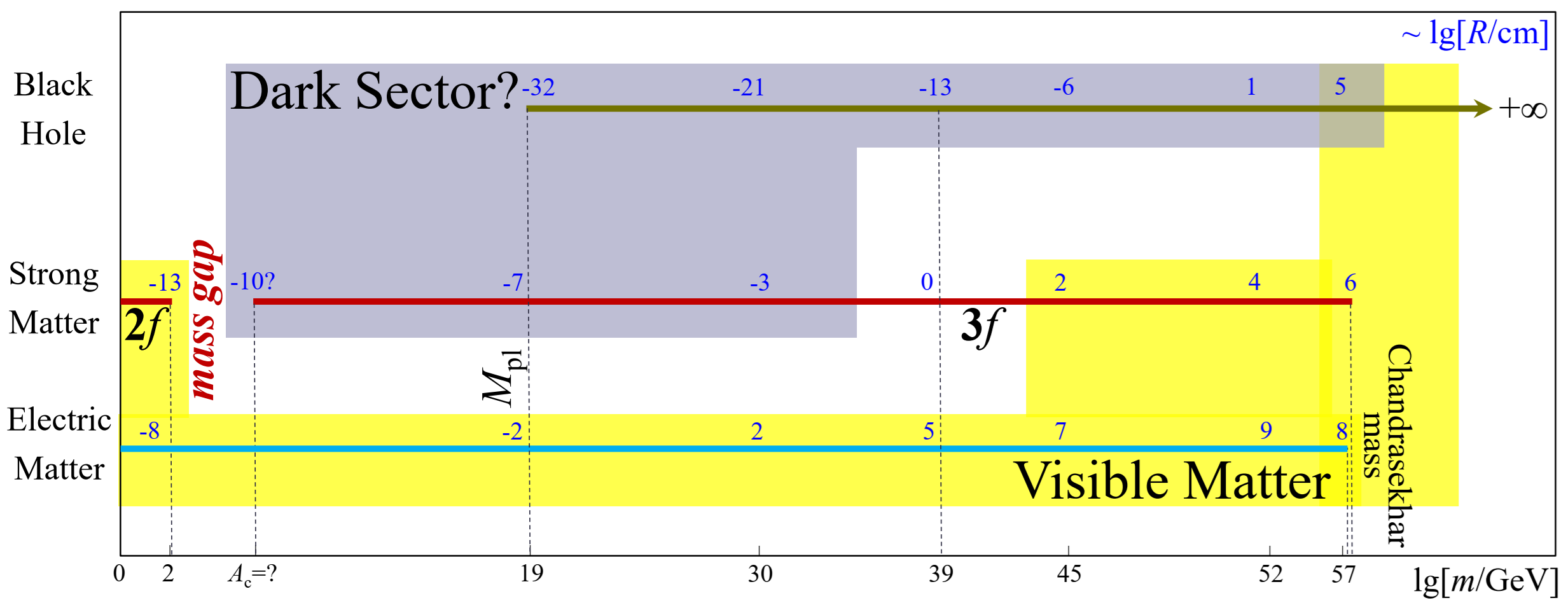}
\caption{The mass spectrum of the black hole, the strong matter, and the electric matter (i.e. the ordinary matter). The minimum mass of the black hole is the Planck mass. The minimum mass of the strong matter depends on the critical baryon number $A_{\mathrm c}$, which is uncertain because of the unknown behavior of the non-perturbative strong interaction. This picture is from~\citep{SnStarBook}.\label{pic:mass}}
\end{figure}

The SnM can also form small nuggets, called strangeon nuggets (SnNs). The baryon number of smallest stable SnN, i.e. the critical baryon number $A_{\mathrm c}$, is uncertain due to the difficulty of computation of the non-perturbative strong interaction. If one considers the role of the electron in breaking the symmetry of $e^\pm$~\citep{critical}, the critical baryon number can be estimated from the Compton wavelength of the electron, $A_{\mathrm c}\approx(\lambda_{\mathrm C}/r_{\rm {nucleon}})^3\approx10^{10}$.
In fact, we have not yet conducted a very detailed calculation on the analysis of weak effects on the stability of strangeon matter, because the strong interaction plays a decisive role in the absolute stability of strange matter in terms of energy. Like the Bodmer-Witten assumption, we assume that the energy of bulk strangeon matter at zero temperature and zero pressure is less than 930 MeV. For little strangeon nuggets with $A<A_{\mathrm c}$, protons and neutrons evaporate from them. At that time, finite volume effects such as surface tension could not be ignored. The evaporation rate is suppressed by the weak interaction because the evaporation requires inverse-strangenization.
We have only ever observed strong interacting matter with baryon number $A<300$ (atomic nuclei) or $A\approx10^{57}$ (pulsars). If this assumption holds, the mass spectrum of strong interacting matter can be broadened considerably~\citep{massspectrum1,massspectrum2}, see Figure~\ref{pic:mass}.

\begin{figure}[H]
\isPreprints{\centering}{} 
\includegraphics[width=10.5 cm]{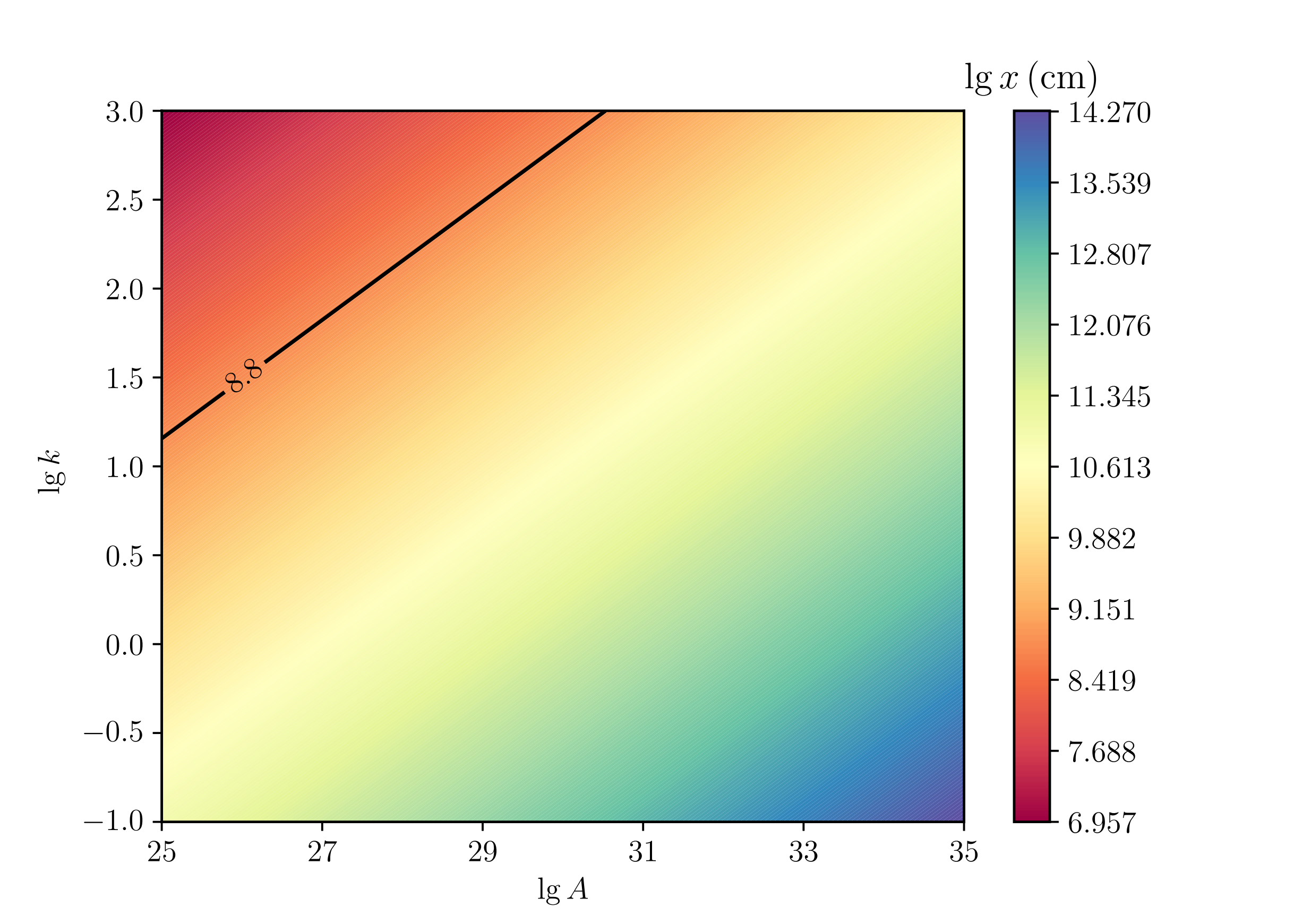}
\caption{The penetration depth $x$ of the strangeon nugget (SnN) in Earth. The black line, $10^{8.8}\,$cm, is the radius of the Earth. The dimensionless $k$ factor is the ratio of the assumed interaction cross-section between the SnN and the ordinary matter and the geometric cross-section of the SnN. This picture is from~\citep{acoustic}.\label{pic:xAk}}
\end{figure}

\begin{figure}[H]
\isPreprints{\centering}{} 
\includegraphics[width=10.5 cm]{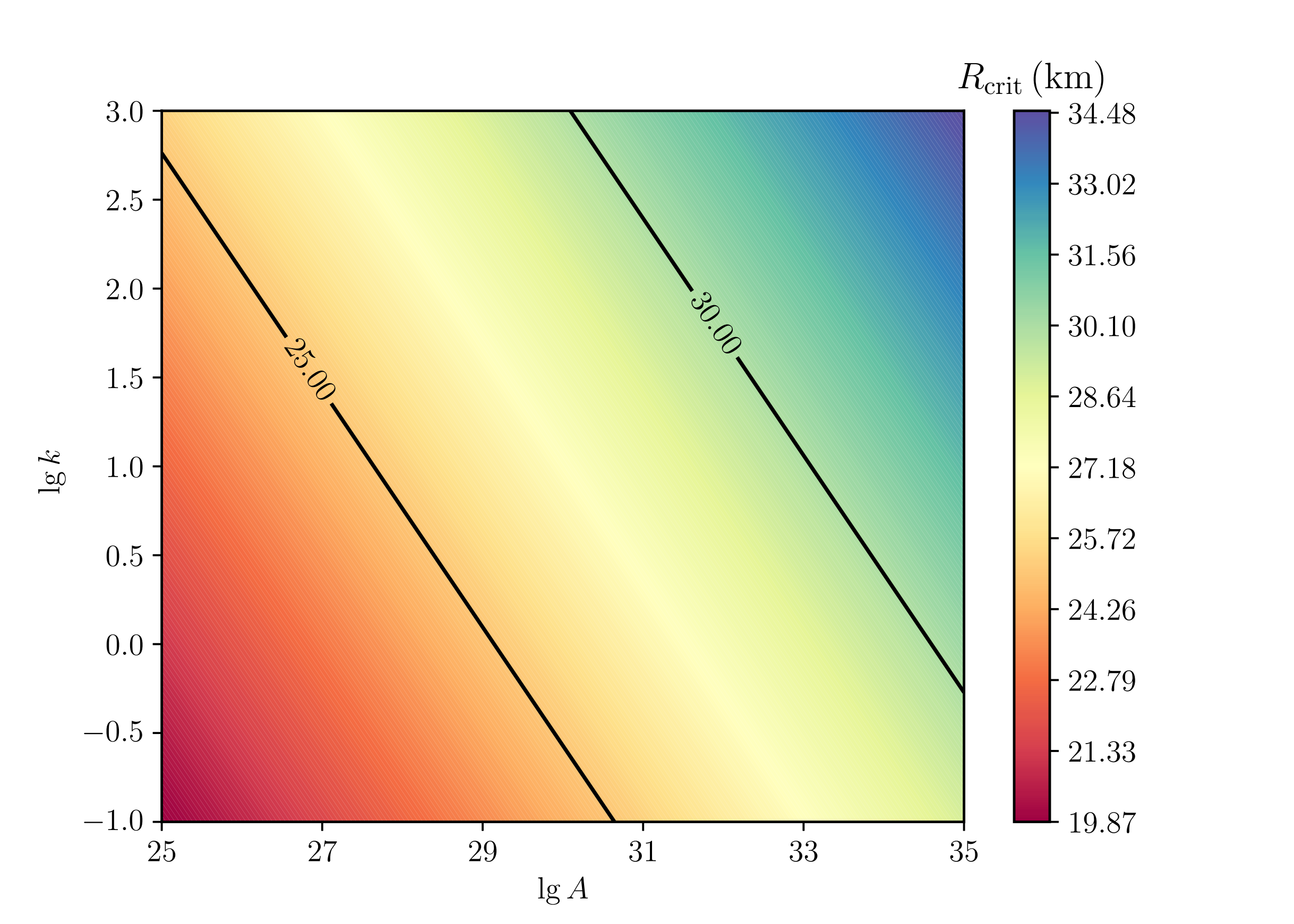}
\caption{The estimated critical detectable distance $R_{\rm crit}$ of the acoustic SnN signal. This picture is from~\citep{acoustic}.\label{pic:CriticalDistance}}
\end{figure}

SnNs with $10^{25}<A<10^{35}$, as quark nuggets, are among the candidates for dark matter. Since the light-three-flavor symmetry in SnM is only slightly broken by the larger strange-quark mass, the charge-to-mass ratio of SnM is low, the electromagnetic signals from SnN are too weak to be observed. SnNs move virially in the universe at a typical speed of about $10^{-3}c$, when they hit Earth, they may not generate a high-energy shower signal, but rather an acoustic shock wave. So, we can detect these non-relativistic nuggets using the acoustic method~\citep{acoustic}. Due to the extremely high density of SnNs, their interaction cross-section with the ordinary matter is small. The cross-section of strangeon nuggets is defined as $\sigma_{\rm SN}=k ~\cdot~(A/10^{30})^{2/3}\cdot 8.0\times 10^{-6}\,{\rm cm^{2}}$, in which $(A/10^{30})^{2/3}\cdot 8.0\times 10^{-6}\,{\rm cm^{2}}$ is the geometric cross-section. Therefore, SnNs' penetration depth in rock or water is very large, even exceeding the radius of the Earth. The calculation results are shown in Figure~\ref{pic:xAk}.
The initial shock wave is strong and sharp but will decay and widen during propagation in the medium due to diffusion, dispersion, dissipation and scattering. The decay can be described as an exponential function, and the frequency drop can be described as a polytropic function of the propagation distance, such as $R^{-1/4}$ or $R^{-1}$, depending on the medium. At a distance about 1\,km, the frequency of the pulse signal is about 0.5\,kHz to 5\,MHz.  At a critical distance $R_{\rm crit}$, the acoustic signal decay to the level of surrounding noise, $R_{\rm crit}$ is the maximum detectable distance. This distance depends on the state, the temperature and the uniformity of the medium. An estimated result is in Figure~\ref{pic:CriticalDistance}, $R_{\rm crit}\simeq30\,$km. 
More accurate results require on-site acoustic experiments.
%

\section{Conclusions} \label{concl}

We proposed that pulsar-like compact stars could be SnSs if Nature favors the three-flavor symmetry of light quarks during core-clollapse supernovae.
With FAST's polarization observation, non-symmetric sparking is found, which could imply ``zits'' (i.e., small hills) on the surface of solid SnS.
Additionally, SnNs with a baryon number of $A\gtrsim 10^{10}$ could also exist, and we could potentially detect them using an acoustic array in the future.
These nuggets, which could manifest in the form of dark matter, may have been produced during the separation of the QCD phases in the early
Universe, setting the initial conditions for big bang nucleosynthesis (BBN).


\vspace{6pt}

\authorcontributions{Conceptualization, Renxin Xu; investigation, Haoyang Qi; writing---original draft preparation, Haoyang Qi; writing---review and editing, Renxin Xu; All authors have read and agreed to the published version of the manuscript.}

\funding{This research received no funding.}


\conflictsofinterest{The authors declare no conflicts of interest.} 



\abbreviations{Abbreviations}{
The following abbreviations are used in this manuscript:
\\

\noindent 
\begin{tabular}{@{}ll}
SnM & Strangeon matter\\
SnN & Strangeon nugget\\
SnS & Strangeon star\\

\end{tabular}
}




\reftitle{References}

\bibliography{reference}

\PublishersNote{}

\end{document}